\documentclass[runningheads]{llncs}

\usepackage[T1]{fontenc}
\usepackage{graphicx}
\usepackage{amsmath}
\usepackage{amssymb}          
\usepackage{booktabs}
\usepackage{multirow}
\usepackage{float}

\usepackage{color}

\graphicspath{{figures/}}

\makeatletter
\renewcommand\section{\@startsection{section}{1}{\z@}%
                       {-13\p@ \@plus -3\p@ \@minus -3\p@}%
                       {8\p@ \@plus 3\p@ \@minus 3\p@}%
                       {\normalfont\large\bfseries\boldmath
                        \rightskip=\z@ \@plus 8em\pretolerance=10000 }}
\renewcommand\subsection{\@startsection{subsection}{2}{\z@}%
                       {-13\p@ \@plus -3\p@ \@minus -3\p@}%
                       {6\p@ \@plus 3\p@ \@minus 3\p@}%
                       {\normalfont\normalsize\bfseries\boldmath
                        \rightskip=\z@ \@plus 8em\pretolerance=10000 }}
\renewcommand\paragraph{\@startsection{paragraph}{4}{\z@}%
                       {-9\p@ \@plus -3\p@ \@minus -3\p@}%
                       {-0.4em \@plus -0.2em \@minus -0.1em}%
                       {\normalfont\normalsize\itshape}}
\makeatother

\begin{document}

%
%
%
\title{Volumetric Latent Diffusion for Brain Tumour Segmentation
       with Missing Modalities}

\titlerunning{Volumetric Latent Diffusion with Missing Modalities}

\author{Hai-Toan Nguyen\inst{1}\thanks{Corresponding author} \and
Tuan-Hung Nguyen\inst{1} \and
Son-Ba Luong\inst{1} \and
Ngoc-Huynh Trinh\inst{1}}

\authorrunning{H.-T. Nguyen et al.}

\institute{Institute for Artificial Intelligence, VNU University of Engineering
and Technology, Hanoi 10000, Vietnam\\
\email{\{nguyenhaitoan, 25025069, bals, huynhtn\}@vnu.edu.vn}
}

\maketitle

\begin{abstract}

    %
    %
    
    Brain tumour segmentation from magnetic resonance imaging normally assumes
    that all four imaging sequences are available, but in clinical practice one or
    more are frequently missing. Existing methods for incomplete inputs learn
    modality-invariant representations or distil from a full-modality teacher, and
    return a single deterministic mask. Generative formulations instead model a
    distribution over plausible masks, but the cost of sampling at full resolution
    has largely confined them to two dimensions. This paper proposes a
    three-dimensional conditional latent diffusion framework for whole-tumour
    segmentation from incomplete multi-sequence MRI. Images and annotations are
    first compressed by two volumetric autoencoders, which makes diffusion over
    whole volumes affordable, and sequences are dropped at random during
    training so that a single model handles every combination of missing
    sequences. Because the model samples rather than predicts, repeated draws
    also yield a voxel-wise map of disagreement at no extra training cost.
    Evaluations on 2,601 studies from BraTS 2023 and 2024 show that the method
    reaches $0.955$ Dice with all four sequences and $0.904$ with FLAIR alone,
    and that it is the most accurate of the four methods compared in eight of
    the fifteen configurations. The disagreement between draws widens as
    sequences are withheld, giving a signal that accompanies the prediction
    rather than replacing it.
    
    \end{abstract}
    
%
%
%

\section{Introduction}
\label{sec:intro}

Segmenting gliomas on magnetic resonance imaging (MRI) supports diagnosis, surgical planning and the assessment of treatment response. The standard protocol acquires four complementary sequences: T1, contrast-enhanced T1 (T1CE), T2 and FLAIR. No single one reveals the whole lesion: enhancement marks the active core, while oedema is clearest on FLAIR. Manual delineation is slow and varies between readers, making automated segmentation a central problem~\cite{brats_menze}.

Deep encoder--decoder architectures~\cite{unet,nnunet} perform well on public benchmarks, but the complete protocol is often unavailable in practice. Methods for incomplete multimodal segmentation~\cite{hemis,rfnet,mmformer,unime} maintain accuracy on partial inputs, yet share an assumption that is rarely stated: the output should be a single deterministic mask. Their evaluations report how accuracy degrades as sequences are removed, but the prediction is returned in the same form either way, giving the clinician no indication of the model's confidence. Generative methods model a distribution over plausible masks rather than one answer~\cite{probunet,phiseg,medsegdiff}, which suits a target whose boundary is genuinely ambiguous. Diffusion formulations are the strongest of these, but each sample costs many network evaluations at full resolution, so they have largely been confined to two dimensions; while brain tumour segmentation is inherently volumetric. Moving the process into a learned latent space removes that cost, and pairing it with modality dropout lets one model serve every acquisition protocol. A sampling formulation also gives predictive uncertainty for free: repeated draws under the same input disagree where the model is unsure, and evaluating each case under every subset makes that comparison paired, so lesion difficulty cancels.

In this paper, we present a three-dimensional conditional latent diffusion framework for whole-tumour segmentation from incomplete multi-sequence MRI, and use it to test this behaviour. Our contributions are:
\begin{enumerate}
    \item \textbf{Volumetric latent diffusion with modality dropout.} We extend latent diffusion segmentation to 3D multi-sequence MRI, training with randomly resampled modality availability so one model serves all
    fifteen configurations rather than one per protocol.

    \item \textbf{Segmentation under every acquisition configuration.} We report whole-tumour accuracy for all fifteen subsets of the four sequences on $2{,}601$ studies from BraTS 2023 and 2024.

    \item \textbf{Predictive uncertainty as a by-product of sampling.} The same repeated draws that form the prediction also give a voxel-wise disagreement map, which we report for every configuration and show to rise
    as sequences are withdrawn.
\end{enumerate}

%
%

\section{Related Work}

\subsection{Incomplete Multimodal Tumour Segmentation}

Clinical acquisitions frequently deviate from the four-sequence protocol that benchmarks assume, and the literature responds in three ways. Synthesis generates the absent sequence and passes a completed set to a conventional network, which leaves the segmentation architecture unchanged but makes accuracy dependent on the quality of the synthesis. Modality-invariant representations avoid that dependence: HeMIS~\cite{hemis} pools per-modality features using statistics that are defined for any subset, and RFNet~\cite{rfnet} adds region-aware fusion that weights sequences differently for different tumour components, since enhancement and oedema are visible in different sequences. Distillation transfers knowledge from a full-modality teacher to a student that observes only a subset; mmFormer~\cite{mmformer} combines modality-specific encoders with inter-modality attention to model long-range dependencies across sequences. UniME~\cite{unime} separates representation learning from fusion, pretraining a shared encoder with masked self-supervision before any fusion module is attached, so that fusion need not simultaneously learn what each sequence contributes and how to combine them. These methods differ in mechanism but agree in output: each returns a single deterministic mask, whatever subset it was given. Accuracy is reported as a function of the available sequences; the confidence attached to the prediction is not. How a model's expressed confidence should respond when evidence is
withdrawn is, to our knowledge, not addressed in this literature.

\subsection{Generative Approaches to Segmentation}

Discriminative networks such as U-Net~\cite{unet} and its successors~\cite{nnunet} learn a mapping that is one-to-one by construction, so the disagreement observed between expert annotators cannot appear in the output. Latent-variable models relax this by decoding samples of a stochastic code into different masks: the Probabilistic U-Net~\cite{probunet} draws a global latent from a conditional prior, and PHiSeg~\cite{phiseg} extends the idea to a hierarchy of latents at several resolutions. Both are limited by the capacity of that latent and by the tendency of the KL term to make the decoder ignore it. Diffusion models~\cite{ddpm} build the distribution through iterative refinement instead of a single stochastic bottleneck. Conditioning a denoising network on the input image turns generation into segmentation, and repeated sampling gives an ensemble from one set of weights; MedSegDiff~\cite{medsegdiff} strengthens the conditioning with frequency-domain filtering of the guidance signal. The cost is many full-resolution network evaluations per sample, which is acceptable in two dimensions and prohibitive in three. Latent diffusion~\cite{ldm} removes that
cost by working in a compressed space, an approach applied to medical segmentation by LSegDiff~\cite{lsegdiff}. MedSegLatDiff~\cite{medseglatdiff} pairs two autoencoders with a latent
diffusion model and shows that a foreground-weighted reconstruction loss is needed when targets are small, while LatentFM~\cite{latentfm} substitutes flow matching to shorten sampling. These methods operate on two-dimensional images with all inputs present, and evaluate on datasets whose ambiguity is fixed by the annotations. The present work extends the line to volumetric data and to inputs that are deliberately incomplete.

%
%

\section{Method}

\begin{figure}[t]
    \centering
    \includegraphics[width=\textwidth]{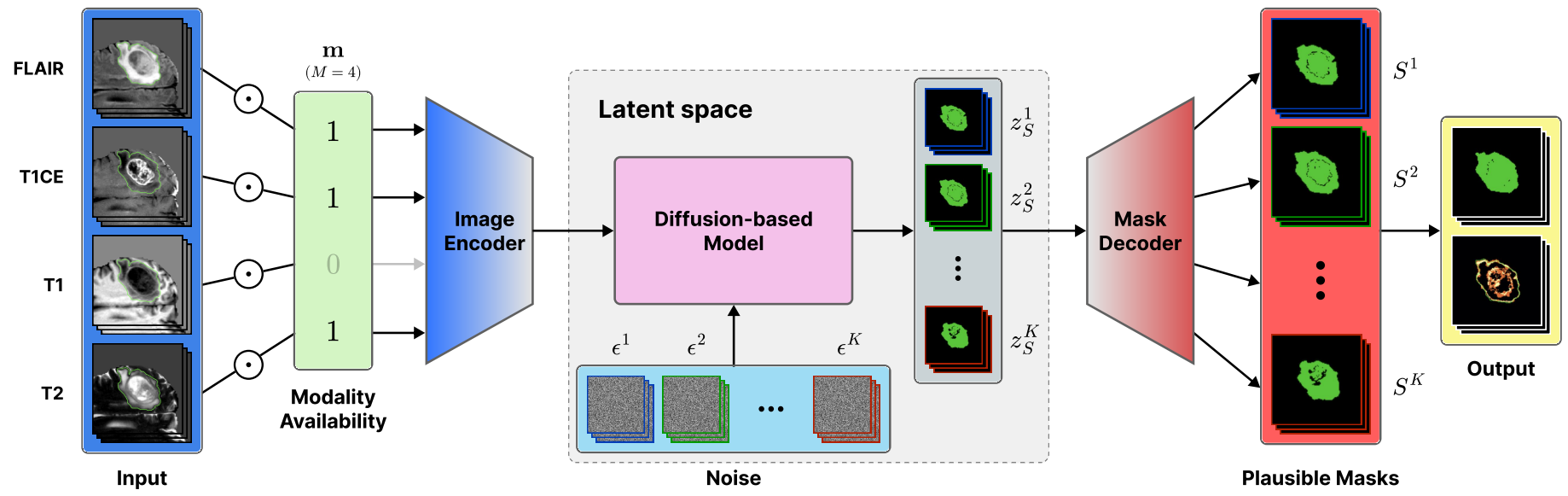}
    \caption{The inference process: The available modalities are encoded once by the pretrained multimodal encoder into a conditioning latent that guides the diffusion reverse process. $K$ independent Gaussian noise samples (e.g., the blue-, green-, and red-bordered noise samples) are denoised in parallel and decoded by the pretrained mask decoder to generate the corresponding $K$ plausible segmentation masks (e.g., the blue-, green-, and red-bordered masks). Their mean is used as the aggregated segmentation, while their per-voxel variance provides the uncertainty map.}
    \label{fig:framework}
\end{figure}

We propose a volumetric latent diffusion framework for segmentation with missing modalities. The model conditions on the available modalities and performs diffusion in a compact latent space, while multiple generated segmentation samples provide an estimate of predictive uncertainty. Figure~\ref{fig:framework} illustrates the overall framework.

\subsection{Problem Formulation}

In this problem setting, we consider a medical image segmentation dataset consisting of paired multimodal observations and voxel-wise annotations. Let
\begin{equation}
    \mathcal{D} = \bigl\{(X^{(i)}, S^{(i)})\bigr\}_{i=1}^{N}
\end{equation}
denote the dataset, where
$X^{(i)} \in \mathbb{R}^{H \times W \times D \times M}$ is a co-registered volume containing $M$ imaging modalities, with $H$, $W$, and $D$ denoting the height, width, and depth in voxels, respectively, and $S^{(i)} \in \{0,1\}^{H \times W \times D}$ is the corresponding voxel-wise expert annotation. We consider binary whole-tumour segmentation, where the target $S$ defines a single foreground region.

\subsubsection{Modality Availability} In routine practice, one or more imaging modalities may be missing due to incomplete acquisition, motion corruption, or differences in acquisition protocols. We explicitly represent the available modalities using a binary vector
\begin{equation}
    \mathbf{m} \in \{0,1\}^{M} \setminus \{\mathbf{0}\},
\end{equation}
where $m_j=1$ indicates that the $j$-th modality is available and $m_j=0$ indicates that it is missing. This gives $2^M-1$ possible non-empty modality configurations. Given an availability vector $\mathbf{m}$, the observed input is defined as
\begin{equation}
    X_{\mathbf{m}} = X \odot \mathbf{m},
\end{equation}
where $\odot$ denotes element-wise multiplication with $\mathbf{m}$ broadcast across the spatial dimensions, such that missing modalities are represented by zero-valued volumes.

\subsubsection{Objective} We formulate segmentation as a probabilistic generative task rather than a deterministic mapping from the observed input to a single segmentation. Following our previous work~\cite{medseglatdiff}, we model the conditional distribution
\begin{equation}
    p(S \mid X),
\end{equation}
which describes the distribution of plausible segmentations given the observed
input $X$. When modalities are missing, the observed input is represented by
$X_{\mathbf{m}}$. We therefore learn a single conditional generative model that
captures the segmentation distribution across different modality
configurations:
\begin{equation}
    p(S \mid X_{\mathbf{m}}).
\end{equation}
At inference time, multiple samples can be generated for the same incomplete
input, whose variation provides an estimate of predictive uncertainty under
different levels of modality missingness.

\subsection{Volumetric Perceptual Compression} To make volumetric diffusion computationally tractable, we perform the generative process in a compact latent space following the latent diffusion paradigm~\cite{ldm}. Both the multimodal input and the segmentation annotation are transformed from the original voxel space into lower-dimensional latent representations. For a volumetric input with spatial dimensions $H \times W \times D$, the corresponding latent representation has spatial dimensions $h \times w \times d$, with the spatial downsampling factor defined as
\begin{equation}
    f = \frac{H}{h} = \frac{W}{w} = \frac{D}{d}.
\end{equation}
The latent representation therefore has the form $z \in \mathbb{R}^{h \times w \times d \times c}$, where $c$ denotes the number of latent channels. This spatial compression substantially reduces the number of spatial positions processed by the diffusion model, making volumetric generation computationally feasible.

\subsubsection{Multimodal Autoencoder} The multimodal encoder $\mathcal{E}_{X}$ maps the incomplete multimodal input $X_{\mathbf{m}}$ into a compact latent representation. To impose a probabilistic structure on the latent space, the encoder parameterizes the latent representation as a Gaussian posterior
\begin{equation}
    q_X(z_X \mid X_{\mathbf{m}}) = \mathcal{N}(\mu_X,\sigma_X^2),
\end{equation}
where $\mu_X$ and $\sigma_X^2$ are the mean and variance predicted by the encoder. We adopt a standard Gaussian prior $p(z_X)=\mathcal{N}(0,I)$ and regularize the posterior toward this prior to obtain a smooth and well-structured latent space for subsequent diffusion modeling. A latent sample $z_X$ is passed through the decoder $\mathcal{G}_{X}$ to reconstruct the input. The autoencoder is trained using the ELBO-based objective
\begin{equation}
    \mathcal{L}_{X} =
    \bigl\|X_{\mathbf{m}}-\mathcal{G}_{X}(z_X)\bigr\|_2^2
    +
    \beta_X D_{\mathrm{KL}}\bigl(q_X(z_X\mid X_{\mathbf{m}})\,\|\,p(z_X)\bigr),
\end{equation}
where the first term is the MSE reconstruction loss that encourages the latent representation to preserve information from the input, while the second term regularizes the posterior toward the Gaussian prior. The coefficient $\beta_X$ controls the trade-off between reconstruction fidelity and latent-space regularization. After training, the posterior mean is used as the multimodal latent, $z_X=\mu_X$, which serves as the conditioning representation for the diffusion model.

\subsubsection{Mask Autoencoder} The mask encoder $\mathcal{E}_{S}$ similarly maps the binary whole-tumour segmentation mask $S$ to a Gaussian posterior
\begin{equation}
    q_S(z_S \mid S) = \mathcal{N}(\mu_S,\sigma_S^2),
\end{equation}
with the same standard Gaussian prior $p(z_S)=\mathcal{N}(0,I)$. A latent sample $z_S$ is decoded by $\mathcal{G}_{S}$ to reconstruct the segmentation mask. Since $S$ is binary with a highly imbalanced foreground and background, we use a weighted binary cross-entropy reconstruction loss instead of the MSE used for the multimodal input. The resulting ELBO-based objective is
\begin{equation}
    \mathcal{L}_{S} =
    -\frac{1}{|V|} \sum_{v\in V} \left[w_{+}S_v\log\hat{S}_v + (1-S_v)\log(1-\hat{S}_v)\right]
    +
    \beta_S D_{\mathrm{KL}} \bigl(q_S(z_S\mid S)\,\|\,p(z_S)\bigr),
\end{equation}
where $\hat{S}=\mathcal{G}_{S}(z_S)$ is the reconstructed mask and $w_{+}>1$ assigns greater importance to foreground voxels. The coefficient $\beta_S$ controls the trade-off between reconstruction fidelity and latent-space regularization. After training, the posterior mean is used as the mask latent, $z_S=\mu_S$, which serves as the target latent for diffusion training. During inference, the pretrained and frozen decoder $\mathcal{G}_{S}$ decodes the generated mask latent into the corresponding segmentation mask.

\subsection{Latent Diffusion with Modality Dropout}

Once the two volumetric perceptual compression models are trained, we perform the generative process in their latent spaces, following the latent diffusion paradigm~\cite{ldm}. Specifically, the diffusion model operates on the segmentation latent $z_S=\mathcal{E}_S(S)$ and learns the conditional distribution $p(z_S\mid z_X)$, where $z_X$ provides the conditioning information derived from the available modalities.

\subsubsection{Forward Diffusion Process} To model the conditional segmentation distribution $p(z_S\mid z_X)$ in the latent space, we adapt a denoising diffusion probabilistic model (DDPM)~\cite{ddpm} to the latent representations. The forward diffusion process starts from the segmentation latent $z_{S,0}=z_S$ and gradually adds Gaussian noise according to
\begin{equation}
    q\bigl(z_{S,t}\mid z_{S,t-1}\bigr)
    =
    \mathcal{N}\bigl(z_{S,t};\sqrt{\alpha_t}\,z_{S,t-1},(1-\alpha_t)I\bigr),
\end{equation}
where $z_{S,t}$ denotes the noisy segmentation latent at timestep $t$, and $\alpha_t$ controls the fraction of signal retained at each diffusion step. After $T$ steps, the distribution of the noisy latent approaches an
isotropic Gaussian distribution. The noisy latent at any timestep can be sampled directly from the segmentation latent as
\begin{equation}
    q\bigl(z_{S,t}\mid z_S\bigr)
    =
    \mathcal{N}\bigl(z_{S,t};\sqrt{\gamma_t}\,z_S,(1-\gamma_t)I\bigr),
    \label{eq:forward-diffusion}
\end{equation}
where $\gamma_t=\prod_{i=1}^{t}\alpha_i$. This formulation avoids explicitly simulating the intermediate diffusion steps during training and allows $z_{S,t}$ to be constructed directly from $z_S$.

\subsubsection{Modality Dropout for Conditional Denoising} During training, we randomly sample a non-empty modality availability vector $\mathbf{m}$ to simulate different missing-modality configurations. Since there are $2^M-1$ possible non-empty configurations, $\mathbf{m}$ is sampled uniformly from this set:
\begin{equation}
    \mathbf{m}\sim\mathcal{U}\left(\{0,1\}^{M}\setminus\{\mathbf{0}\}\right).
\end{equation}
The resulting incomplete input $X_{\mathbf{m}}$ is encoded by the pretrained multimodal encoder to obtain the conditioning latent $z_X=\mathcal{E}_{X}(X_{\mathbf{m}})$. The noisy segmentation latent is then conditioned on $z_X$ through channel-wise concatenation:
\begin{equation}
    z_{S,t}^{\mathrm{cond}} = z_{S,t}\oplus z_X,
\end{equation}
where $\oplus$ denotes channel-wise concatenation. Since $z_S$ and $z_X$ share the same spatial dimensions, $z_X$ can be directly concatenated with $z_{S,t}$ along the channel dimension and provided as the input to the denoising network. The denoising network $\epsilon_{\theta}$ is implemented as a 3D U-Net parameterized by $\theta$ and is trained to predict the Gaussian noise added to the segmentation latent using the standard MSE objective:
\begin{equation}
    \mathcal{L}_{\mathrm{diff}}=
    \mathbb{E}_{(X,S)\sim\mathcal{D},\,\mathbf{m},\,\epsilon,\,t}
    \left[\left\|\epsilon - \epsilon_{\theta} \bigl(z_{S,t}^{\mathrm{cond}},t\bigr) \right\|_2^2 \right].
    \label{eq:diffusion-loss}
\end{equation}
Here, $\mathbf{m}$ is uniformly sampled from all non-empty modality configurations, $t\sim\mathcal{U}\{1,\ldots,T\}$, and $\epsilon\sim\mathcal{N}(0,I)$ denotes the Gaussian noise used to construct $z_{S,t}$ according to Eq.~\eqref{eq:forward-diffusion}.

\subsubsection{Reverse Diffusion Process} At inference time, we follow DDIM sampling~\cite{ddim} to generate the denoised segmentation latent conditioned on the multimodal latent $z_X$. Rather than using all diffusion timesteps, DDIM performs the reverse process on a smaller subset of timesteps, reducing the number of denoising updates. Specifically, the reverse process performs $l$ updates on a subsequence $\{\tau_i\}_{i=0}^{l}$, where $\tau_0=0$ and $\tau_l=T$. For a uniform sampling stride $\Delta_t$, the selected timesteps are given by
\begin{equation}
    \tau_i = i\Delta_t,\quad i=0,1,\dots,l.
\end{equation}
The reverse process starts from a Gaussian noise sample $z_{S,\tau_l}\sim\mathcal{N}(0,I)$ and proceeds from $\tau_l$ to $\tau_0$. At each timestep $\tau_i$, the denoising network $\epsilon_{\theta}$ predicts the noise component from the conditional latent $z_{S,\tau_i}^{\mathrm{cond}}$, which is then used to compute the latent at the previous timestep:
\begin{equation}
z_{S,\tau_{i-1}} =
\sqrt{\gamma_{\tau_{i-1}}},z_{S,0}^{\mathrm{pred}}
+ \sqrt{1-\gamma_{\tau_{i-1}}},
\epsilon_{\theta}\bigl(z_{S,\tau_i}^{\mathrm{cond}},\tau_i\bigr),
\end{equation}
where $i=l,l-1,\dots,1$. Here, the clean latent $z_{S,0}^{\mathrm{pred}}$ is obtained by inverting Eq.~\eqref{eq:forward-diffusion} using the predicted noise. The final denoised segmentation latent is obtained as $\hat{z}_{S,0}$ at $\tau_0$. Finally, the pretrained and frozen mask decoder then produces
$\hat{S}=\mathcal{G}_{S}(\hat{z}_{S,0})$.

\subsection{Sampling and Predictive Uncertainty}

To capture the variability of the predicted segmentation for a given incomplete input, we generate multiple segmentation samples from different initial noise realizations. Specifically, the conditioning latent $z_X$ is shared across all samples, while $K$ independent Gaussian noise samples are drawn as
\begin{equation}
    \epsilon^{i} \sim \mathcal{N}(0,I), \qquad i=1,\ldots,K.
\end{equation}
Each noise sample is independently propagated through the reverse diffusion process, producing a segmentation latent $z_{S}^{i}$. The pretrained and frozen mask decoder then maps each latent to a segmentation $S^{i} = \mathcal{G}_{S}(z_{S}^{i})$. The final segmentation is obtained from the voxel-wise mean of the generated samples, while their variability is quantified using the voxel-wise variance. The final binary segmentation is then obtained by thresholding the mean prediction at $0.5$.

%
%
%
%

\section{Experiments}
\label{sec:experiments}

\subsection{Dataset and Preprocessing}
\label{subsec:dataset}

We evaluate on the BraTS 2023 and 2024 adult glioma
collections~\cite{brats_menze,brats_bakas}, which provide $1{,}251$ and $1{,}350$ studies with publicly released annotations, giving $2{,}601$ in total. Each study provides four co-registered, skull-stripped sequences (T1, T1CE, T2 and FLAIR) at an original resolution of $240 \times 240 \times 155$ voxels, with a voxel-wise expert annotation. Intensities are normalised per sequence and per study by $z$-scoring over brain voxels only, so that background air does not bias the statistics and scanner-dependent ranges are placed on a common scale. Each study is then cropped to a fixed cubic window centred on the centroid of the annotation bounding box. This crop defines the input size, $H = W = D = 128$, and hence a latent grid of $h = w = d = 32$. A fixed crop allows annotation latents to be computed once and cached, and this window retains the complete tumour for $97\%$ of studies. Studies are partitioned patient-wise into training, validation and test sets in a $70{:}20{:}10$ ratio with a fixed seed, giving $1{,}821$, $520$ and $260$ studies respectively. All configurations of a given study fall in the same partition, so no case is ever evaluated under one subset of sequences while having been trained on under another.

\subsection{Implementation Details}
\label{subsec:implementation}

\subsubsection{Autoencoders} Both use a residual convolutional encoder--decoder with two residual units per
level and two downsampling stages, mapping $H \times W \times D$ inputs
to $h \times w \times d$ latents. The image autoencoder uses channel widths $(64, 128, 256)$ and
$\beta_{X} = 10^{-4}$; the annotation autoencoder uses $(32, 64, 128)$,
$C_{S} = 4$ and $\beta_{S} = 10^{-2}$. The annotation autoencoder is trained on
the single binary whole-tumour mask, with
$w_{+} = 4.4$. Both are trained for 200 epochs with
AdamW~\cite{adam} at learning rate $10^{-4}$, weight decay $10^{-4}$ and batch
size 32, with early stopping after 25 validation checks without improvement,
then frozen. Modality dropout is applied to the image autoencoder as well as
the diffusion model, so its encoder is not shown partial inputs for the first
time at inference.

\subsubsection{Diffusion model}
The denoising network is a 3D U-Net with base width 64, channel multipliers $(1, 2, 4)$ and two residual blocks per level, totalling $31.3$M parameters. The image latent is projected from 256 to 64 channels before concatenation with the noisy annotation latent. Training uses $T = 1000$ timesteps with a linear schedule, $2 \times 10^{5}$ steps, batch size 64 and AdamW at learning rate $10^{-4}$ and weight decay $10^{-4}$, with $\mathbf{m}$ resampled uniformly over the 15 non-empty configurations for every example.

\subsubsection{Inference} Sampling uses DDIM~\cite{ddim} with 200 steps, beyond which accuracy changes little while cost grows in proportion. We draw $K = 5$ samples per case, average them, and threshold at $\tau = 0.5$; Figure~\ref{fig:n-sweep} justifies that choice. All experiments use a fixed seed and are implemented in PyTorch; training was performed on NVIDIA H200 GPUs. 

\subsubsection{Evaluation protocol} Every test case is evaluated under all 15 modality configurations, giving a complete case $\times$ configuration grid. The whole tumour is the only target: we do not segment or report the tumour core or the enhancing tumour. Because the autoencoder is frozen, its reconstruction fidelity upper-bounds what any diffusion model in that latent space can reach, so Table~\ref{tab:autoencoder} reports it separately and Table~\ref{tab:modality} should be read against that ceiling.

\subsubsection{Metrics} For anomaly segmentation, we use the Dice coefficient and Intersection over Union (IoU), which measure the overlap between the predicted and ground-truth masks. Dice is used as the primary segmentation metric. For reconstructed image quality, we use the structural similarity index (SSIM) and peak signal-to-noise ratio (PSNR), which assess the similarity and fidelity of reconstructed images to their corresponding reference images. All four metrics are higher-is-better measures.

\subsection{Results and Discussion}

\begin{table}[b!]
    \centering
    \caption{Reconstruction fidelity of the frozen autoencoders, which bounds
    what any diffusion model in these latent spaces can attain. Dice and IoU
    measure mask overlap; SSIM and PSNR measure image similarity. Higher is
    better for all four. The image autoencoder is reported per sequence.}
    \label{tab:autoencoder}
    \begin{tabular}{llcc@{\hspace{2em}}c@{\hspace{2em}}c}
        \toprule
        Autoencoder & Target & Dice & IoU & SSIM & PSNR (dB) \\
        \midrule
        Annotation & WT     & 0.9729 & 0.9475 & 0.8833 & 24.87 \\
        \midrule
        \multirow{4}{*}{Image}
                   & FLAIR  & n/a & n/a & 0.8687 & 23.55 \\
                   & T1CE   & n/a & n/a & 0.8527 & 19.62 \\
                   & T1     & n/a & n/a & 0.8977 & 22.73 \\
                   & T2     & n/a & n/a & 0.9374 & 24.59 \\
        \bottomrule
    \end{tabular}
\end{table}

The frozen annotation autoencoder reconstructs the whole tumour at a Dice of $0.973$ (Table~\ref{tab:autoencoder}), which is the ceiling every segmentation number below is measured against; the remaining gap is attributable to the diffusion model rather than to compression. Among the four sequences the image autoencoder reconstructs T1CE least well, at an SSIM of $0.853$ against $0.937$ for T2, and T1CE is also the sequence that contributes least to the whole tumour, so the two limitations compound rather than offset.

\begin{figure}[t!]
    \centering
    \includegraphics[width=0.62\textwidth]{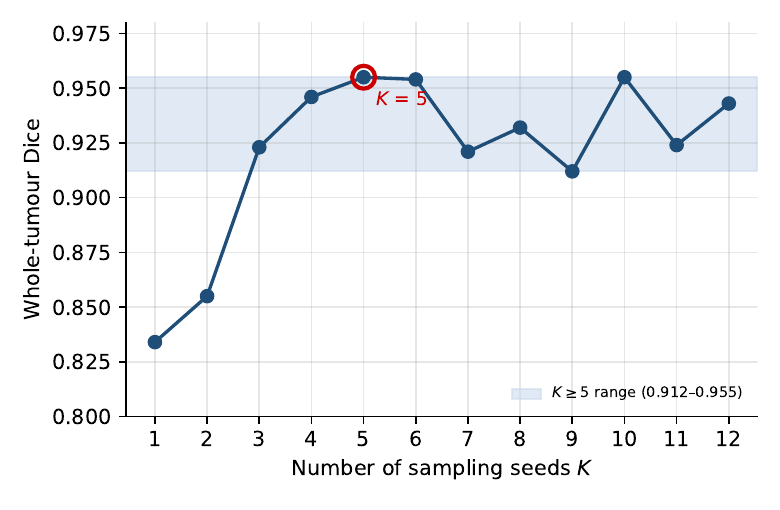}
    \caption{Whole-tumour Dice against the number of sampling seeds $K$, with
    all four sequences available. Accuracy rises sharply to $K = 5$ and shows
    no further trend beyond it; the shaded band is the range spanned by
    $K \geq 5$. We use $K = 5$, the smallest ensemble reaching the maximum.}
    \label{fig:n-sweep}
\end{figure}

Figure~\ref{fig:n-sweep} shows whole-tumour Dice against the number of sampling seeds on the fully observed configuration. Accuracy rises sharply from $0.834$ at $K = 1$ to $0.955$ at $K = 5$ and shows no further trend beyond that point, fluctuating between $0.912$ and $0.955$ over $5 \leq K \leq 12$. We therefore use $K = 5$, the smallest ensemble that attains the maximum. The residual fluctuation is worth stating plainly: at roughly $\pm 0.02$ Dice it is of the same order as the margins separating the methods in Table~\ref{tab:modality}, so individual configurations should be read as indicative and the comparison rests on the pattern across all fifteen rather
than on any single cell.

\begin{table}[t!]
    \centering
    \caption{Whole-tumour Dice across all 15 modality configurations. All
    methods are evaluated on the same split and receive identical crops, so
    the columns are directly comparable; because those crops are centred on
    the annotation (Section~\ref{subsec:limitations}), the absolute values are
    not comparable with whole-volume results in the literature. Best result
    per configuration in bold.}
    \label{tab:modality}
    \setlength{\tabcolsep}{4pt}
    \small
    \begin{tabular}{llcccc}
        \toprule
        $|\mathbf{m}|$ & Available sequences
            & Ours & HeMIS & RFNet & UniME \\
        \midrule
        4 & FLAIR+T1CE+T1+T2 & \textbf{0.955} & 0.821 & 0.901 & 0.933 \\
        \midrule
        \multirow{4}{*}{3}
          & FLAIR+T1CE+T1  & 0.889 & 0.767 & 0.880 & \textbf{0.891} \\
          & FLAIR+T1CE+T2  & \textbf{0.922} & 0.832 & 0.880 & 0.917 \\
          & FLAIR+T1+T2    & \textbf{0.914} & 0.863 & 0.902 & 0.911 \\
          & T1CE+T1+T2     & 0.766 & 0.702 & \textbf{0.777} & 0.765 \\
        \midrule
        \multirow{6}{*}{2}
          & FLAIR+T1CE     & 0.890 & 0.800 & 0.899 & \textbf{0.900} \\
          & FLAIR+T1       & 0.880 & 0.782 & 0.812 & \textbf{0.902} \\
          & FLAIR+T2       & \textbf{0.919} & 0.812 & 0.857 & 0.912 \\
          & T1CE+T1        & \textbf{0.646} & 0.341 & 0.619 & 0.635 \\
          & T1CE+T2        & \textbf{0.630} & 0.582 & 0.593 & 0.627 \\
          & T1+T2          & 0.681 & 0.582 & 0.509 & \textbf{0.702} \\
        \midrule
        \multirow{4}{*}{1}
          & FLAIR          & \textbf{0.904} & 0.701 & 0.827 & 0.897 \\
          & T1CE           & 0.281 & 0.223 & 0.324 & \textbf{0.329} \\
          & T1             & 0.355 & 0.192 & 0.404 & \textbf{0.423} \\
          & T2             & \textbf{0.578} & 0.251 & 0.426 & 0.562 \\
        \midrule
        \multicolumn{2}{l}{Mean over all 15}
                           & 0.747 & 0.617 & 0.707 & \textbf{0.754} \\
        \bottomrule
    \end{tabular}
\end{table}

Table~\ref{tab:modality} reports whole-tumour Dice for every configuration against three incomplete-modality baselines. With all four sequences the proposed model reaches $0.955$, above UniME at $0.933$, RFNet at $0.901$ and HeMIS at $0.821$, and it is the strongest of the four in eight of the fifteen configurations. Averaged over all fifteen, however, it reaches $0.747$ against $0.754$ for UniME, $0.707$ for RFNet and $0.617$ for HeMIS: the proposed model leads wherever FLAIR is present and in the fully observed case, while UniME is stronger on the sparse T1CE- and T1-dominated subsets, so the two are separated by less than the spread across configurations. The ordering across configurations is not governed by how many sequences are supplied. FLAIR alone attains $0.904$, within $0.051$ of the fully observed result and above every configuration that omits it; the three sequences other than FLAIR together reach only $0.766$, and T1CE alone reaches $0.281$. This is expected rather than anomalous: the whole tumour is delimited by peritumoural oedema, which is what FLAIR is acquired to display~\cite{brats_menze,brats_bakas}, so most of the information relevant to this target is already present in that one sequence. T1CE instead highlights the enhancing rim and carries little information about the outer boundary being scored. The same asymmetry is reported throughout the missing-modality literature~\cite{hemis,rfnet,mmformer}. The configurations where the proposed model trails the baselines are precisely those without FLAIR, which is also where the compression ceiling and the weaker T1CE reconstruction are most exposed.

\begin{figure}[b!]
    \centering
    \includegraphics[width=\textwidth]{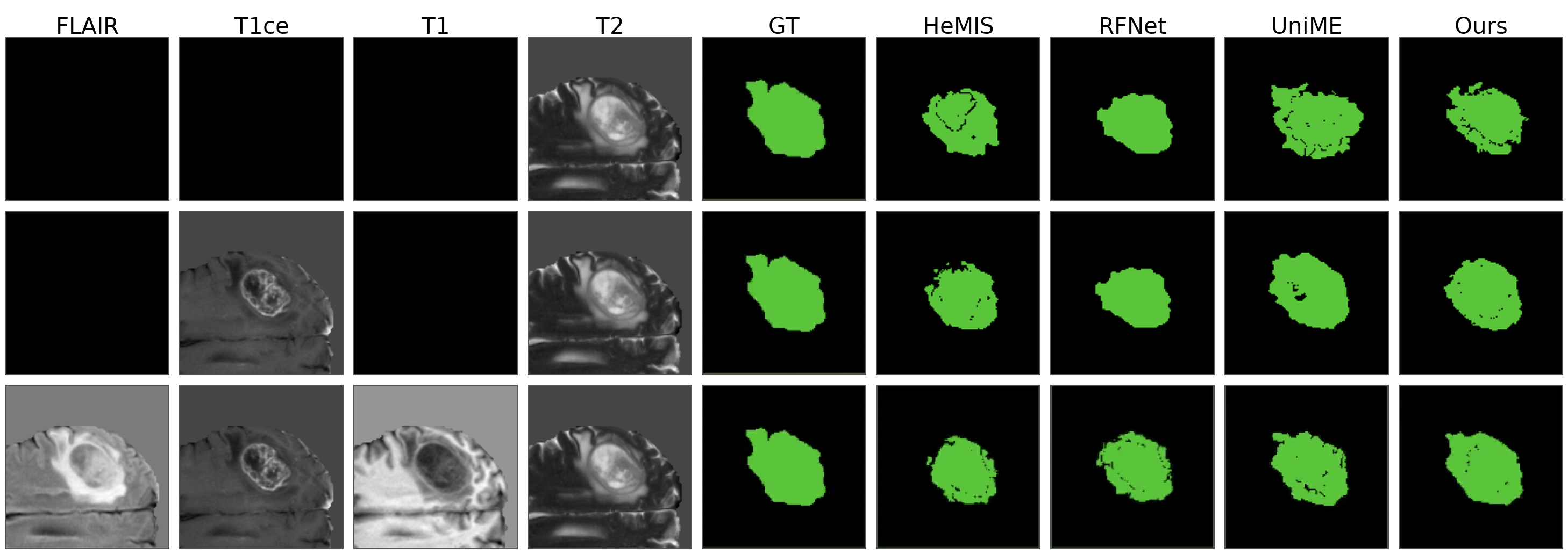}
    \caption{Whole-tumour predictions under three acquisition configurations
    on one case: T2 alone (top), T1CE+T2 (middle) and all four sequences
    (bottom). The first four columns show the sequences supplied, with
    withheld ones left black; the remainder show the annotation (GT), the
    three baselines and the proposed model. With T2 alone the four methods
    differ little on this slice; as sequences are added the proposed model's
    mask tightens towards the annotation, recovering the upper-left extension
    that HeMIS and RFNet miss and leaving fewer interior holes than UniME.}
    \label{fig:qualitative}
\end{figure}

Figure~\ref{fig:qualitative} shows what these differences look like on a single case under three configurations. With T2 alone the four methods produce visibly similar masks on this slice. As T1CE and then FLAIR are added, the proposed model's mask tightens towards the annotation, while HeMIS changes comparatively little; with all four sequences it recovers the upper-left extension that HeMIS and RFNet miss and fragments the interior less than UniME. The ranking in Table~\ref{tab:modality} is over the whole test set.

\begin{figure}[b!]
    \centering
    \includegraphics[width=\textwidth]{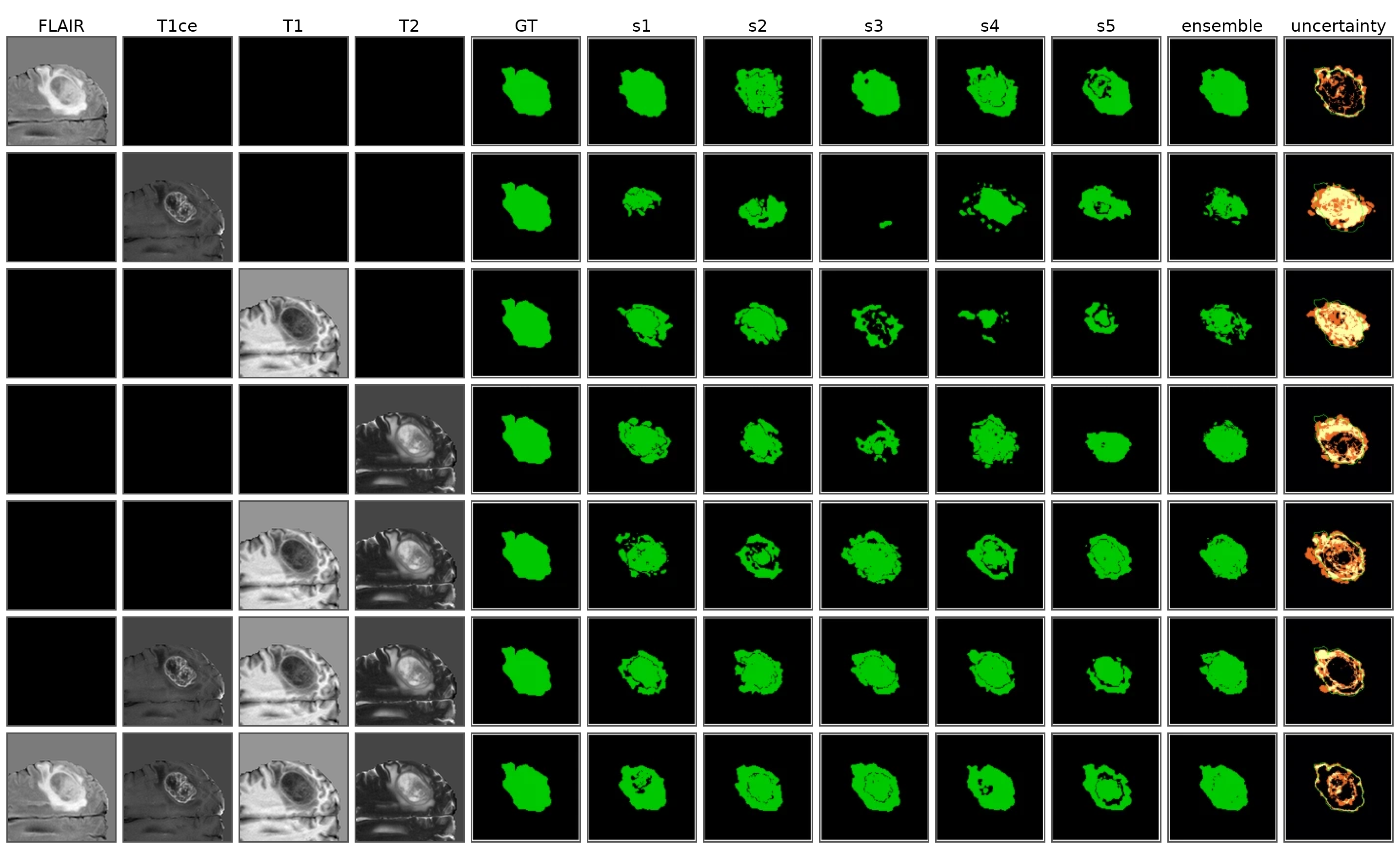}
    \caption{One test case under seven acquisition configurations, shown on a
    single axial slice through the volume; the model and all reported metrics
    operate on the full $128^{3}$ volume. The first
    four columns show the sequences given to the model, with withheld ones
    left black; the remainder show the annotation, five independent sampling
    seeds, their mean, and their per-voxel variance. Rows one to four isolate
    each sequence in turn. Sample disagreement is visibly wider when FLAIR is
    absent, and the uncertainty maps are correspondingly
    brighter and more diffuse.}
    \label{fig:combo-grid}
\end{figure}

Figure~\ref{fig:combo-grid} shows, for the proposed model, how the prediction changes as sequences are withheld. The five sampling seeds agree closely when the input is informative and diverge as it is impoverished: with all four sequences they differ only along a thin band at the tumour boundary, whereas with T1CE alone they disagree over most of the lesion and two of the five recover almost none of it. The uncertainty maps follow the same progression, from a narrow rim to a broad region covering the interior. Every row is the same case and the same annotation, so the only variable is how much of the input the model sees. The figure establishes the mechanism; Table~\ref{tab:modality} carries the claim across the test set.

%

\section{Conclusion}
\label{sec:conclusion}

We presented a three-dimensional conditional latent diffusion framework for whole-tumour segmentation from incomplete multi-sequence MRI. Two volumetric autoencoders compress images and annotations by a factor of four per axis, which makes diffusion over whole volumes tractable, and modality dropout during training lets a single model serve all fifteen acquisition configurations. Repeated sampling yields both a consensus segmentation and a voxel-wise map of disagreement between plausible annotations. Beyond accuracy, we treated modality availability as an experimental variable rather than a nuisance. Withholding a sequence is a controlled reduction in the evidence available, and evaluating every case under every subset makes the resulting change in uncertainty measurable within each case. Our results indicate that the predicted uncertainty follows this behaviour, which suggests it reflects genuine information loss rather than sampling noise. 

\subsection{Limitations and Future Work}
\label{subsec:limitations}

Two aspects of the protocol limit how the numbers should be read. First, the
$128^{3}$ window is centred on the centroid of the annotation bounding box in
every split, so localisation is given rather than inferred and the absolute
Dice values are correspondingly optimistic: they measure segmentation inside a
window already known to contain the lesion, and are not comparable with
results obtained by whole-volume inference. The comparison between methods is
unaffected, since all four receive identical crops. A deployable system would
derive the window from a brain mask, an intensity heuristic or a separate
detector, or dispense with it through sliding-window inference; we have not
re-run the study under that protocol. Second, the number of sampling seeds
$K$ was read off Figure~\ref{fig:n-sweep}, which was computed on the test
split rather than on held-out validation data; the remaining hyperparameters
were fixed a priori or selected by early stopping on the validation split,
and any $K \geq 5$ changes Dice by at most $0.043$.

The autoencoders apply a fixed $4\times$ downsampling per axis, chosen so that
diffusion over whole volumes fits within memory, and the diffusion model works
entirely inside that latent space, so the reconstruction fidelity of
Table~\ref{tab:autoencoder} is a ceiling it cannot exceed. We segment the
whole tumour only; extending to the tumour core and the enhancing tumour is
the natural next step, and because those structures are smaller and thinner it
would likely depend more on the compression scheme than on the diffusion
model, which would require repeating the autoencoder study at several
downsampling factors. The annotation autoencoder is trained with a weighted
cross-entropy alone; region-overlap losses such as soft Dice, or
boundary-aware and focal variants, may preserve small structures better at the
same compression factor, and we have not compared them. Finally, the
predictive uncertainty is reported but not compared against other estimators,
so Monte Carlo dropout, deep ensembles and the Probabilistic U-Net would place
it on a common footing with established methods.

\subsubsection{Code availability.}
Training, evaluation and figure code is available at\\
\url{https://github.com/Toan-Nguyen26/medseglatdiff-3d}.


\bibliographystyle{splncs04}
\bibliography{refs}

\end{document}